\documentclass{article}

\usepackage[sglblindworkshop, final]{neurips_2026}
\workshoptitle{Sim2Science}

\usepackage[utf8]{inputenc}
\usepackage[T1]{fontenc}
\usepackage{hyperref}
\usepackage{url}
\usepackage{booktabs}
\usepackage{amsmath}
\usepackage{amsfonts}
\usepackage{nicefrac}
\usepackage{microtype}
\usepackage{xcolor}
\usepackage{graphicx}
\usepackage{multirow}

\title{Reconstruction of Unobserved Core Electron Density in Tokamak
  Plasmas with a Neural Operator Transport Surrogate and
  Uncertainty Quantification}

\author{%
  Anonymous\\
}

\author{%
  D.~Sorokin$^{1}$\thanks{Corresponding author: \texttt{ds@nextfusion.org}} \quad
  A.~Granovskiy$^{1}$ \quad
  M.~Stokolesov$^{1}$ \quad
  E.~Adishchev$^{1}$ \\
  \textbf{M.~Nurgaliev}$^{1}$ \quad
  \textbf{G.~Subbotin}$^{1}$ \quad
  $^{1}$Next Step Fusion, Bertrange, L-8070, Luxembourg \\
}

\begin{document}

\maketitle

\begin{abstract}

Real-time control of tokamak plasmas requires the electron density
profile. However, in a reactor-relevant device, diagnostic port
access is limited and the profile can be observed only partially.
In the present work we reconstruct the density profile by inferring
the coefficients of a physics-based one-dimensional particle
transport equation, using a Fourier Neural
Operator (FNO) surrogate of the transport solver as a fast differentiable forward model.
The surrogate enables realtime-compatible gradient-based reconstruction
and uncertainty quantification from the reflectometry and
interferometry measurements. We compare four uncertainty quantification methods
--- multi-seed ensembles, the Laplace approximation, Metropolis--Hastings,
and a Gaussian method independent of the reconstruction fit. The resulting uncertainty is
scored with CRPS and coverage calibration, separately on the observed and unobserved
regions. We further demonstrate advantage of our approach over standard analytical function fit 
on synthetic and experimental DIII-D discharges.

\end{abstract}

\section{Introduction}
\label{sec:intro}

Tokamak plasma operation relies on knowledge of the radial
electron density profile $n_e(\rho, t)$, where
$\rho \in [0, 1]$ is the normalised radius. In a future fusion power plant, however, the
available density diagnostics can constrain the profile only partially.
This restriction is not incidental: the future
fusion power plant must minimize the number of dedicated diagnostic
ports in a neutron-shielded, tritium-breeding vessel, which rules out
multi-view techniques such as Thomson scattering or multiple interferometer chords. Therefore we restrict
the observation set to a single central interferometer chord (line-averaged
density) and reflectometry (profile outside the pedestal top), both of
which need minimal port access.

Standard practice in profile reconstruction is to fit a parametric pedestal shape
(modified hyperbolic tangent, mtanh~\cite{groebner1998}) to each time
slice independently, which extrapolates into the core without any
physical constraint. However, the density profile evolution must obey a
particle transport equation with diffusivity, pinch velocity, and
particle source coefficients. In this work, we infer these coefficients from
the observations and evolve the equation forward to reconstruct the full
profile, with uncertainty inherited from the posterior distribution over the coefficients.
This is the integrated-data-analysis viewpoint~\cite{fischer2010}, made
tractable here by replacing the PDE solver with a Fourier Neural
Operator~\cite{li2021fno} surrogate: a single forward model evaluation
becomes a GPU-friendly network pass, gradients and Hessians come from
automatic differentiation, and posterior sampling requiring $10^3$--$10^4$
forward evaluations becomes feasible. The approach is complementary to
differentiable transport solvers such as TORAX~\cite{citrin2024torax} and
to control-oriented transport modelling~\cite{felici2012, blanken2018}.

Our contributions are:
(i)~a FNO surrogate of the particle transport equation
with realistic flux-surface geometry, used as the forward model for
Bayesian inference of transport and source parameters from
control-relevant observations (line-averaged density plus edge
profile);
(ii)~a systematic comparison of four posterior approximations, scored
with proper scoring rules and coverage separately on observed and
unobserved regions;
(iii)~benchmarks against classical mtanh approach on synthetic and experimental data.

\section{Problem setup}
\label{sec:setup}

\paragraph{Transport model.}
The electron density $n_e$ evolves on the normalised radius $\rho \in [0,1]$
according to the flux-surface-averaged continuity equation~\cite{astra}
\begin{equation}
  \frac{\partial n_e}{\partial t}
  = \frac{1}{V'}\frac{\partial}{\partial \rho}
    \left[ V' \left( G_1 D \frac{\partial n_e}{\partial \rho}
    - G_0 v\, n_e \right) \right]
  + S_{\mathrm{wall}}(\rho) + S_{\mathrm{core}}(\rho),
  \label{eq:transport}
\end{equation}
where $V'(\rho)$ is the flux-surface volume derivative and
$G_0 = \langle |\nabla\rho| \rangle$,
$G_1 = \langle (\nabla\rho)^2 \rangle$ are geometric coefficients
computed from the magnetic equilibrium (EFIT~\cite{efit} for DIII-D discharges).
Boundary conditions are zero flux at the axis and fixed density at the
separatrix. The diffusivity $D(\rho) \ge 0$ and pinch velocity $v(\rho)$
are piecewise-linear with a single interior breakpoint
$\rho_b$~\cite{blanken2018}; the source is a superposition of an
exponential wall-fuelling term with amplitude $A_w$ and decay length
$\lambda$, and a Gaussian core term with amplitude $A_c$, center
$\rho_0$, and width $\sigma_0$. The resulting parameter vector
$\kappa \in \mathbb{R}^{11}$ collects six transport and five source
parameters, with a uniform box prior $\kappa \in [\kappa_l, \kappa_u]$
set by physically admissible ranges.

\paragraph{Observation model.}
At each output time $t_k$, the observation operators are the line-averaged
density and the edge profile:
\begin{equation}
  y_k^{\mathrm{avg}} = \int_0^1 n_e(\rho, t_k)\, w(\rho)\, d\rho,
  \qquad
  y_k^{\mathrm{edge}} = n_e(\rho_j,\, t_k), \quad \rho_j \ge \rho_{\mathrm{obs}} = 0.7,
  \label{eq:obsop}
\end{equation}
where the line average is along a horizontal midplane interferometer chord
with $w(\rho)$ proportional to the chord-length interval between
consecutive flux ($\psi$) surfaces. The core region $\rho <
\rho_{\mathrm{obs}}$ is unobserved and is reconstructed only through the
transport constraint~\eqref{eq:transport} and the interferometer
observation.

\paragraph{Noise model.}
The following noise model corrupts~\eqref{eq:obsop} for synthetic training
and evaluation trajectories:
\begin{equation}
  \tilde y^{\mathrm{avg}}_k = y_k^{\mathrm{avg}} (1 + \varepsilon_k),
  \qquad
  \tilde y^{\mathrm{edge}}_{k,j} = n_e(\rho_j - \delta_k,\, t_k).
  \label{eq:noisemodel}
\end{equation}
The dominant interferometer error is a per-discharge \emph{calibration}
error, a multiplicative scale factor
$\varepsilon_k \sim \mathcal{N}(0, \sigma_{\mathrm{ifo}}^2)$ shared
across $t_k$ within a discharge. The reflectometry error is dominated by a
radial \emph{position} shift, modelled as a rigid two-component shift:
\begin{equation}
  \delta_k = \delta_{\mathrm{bias}} + \delta_{\mathrm{jit},k},
  \qquad
  \delta_{\mathrm{bias}} \sim \mathcal{N}(0, \sigma_{\mathrm{bias}}^2)
  \;\text{(per discharge)},
  \qquad
  \delta_{\mathrm{jit},k} \sim \mathcal{N}(0, \sigma_{\mathrm{jit}}^2)
  \;\text{(i.i.d.\ per step)},
  \label{eq:posnoise}
\end{equation}
where $\delta_{\mathrm{bias}}$ stems from toroidal-field (density cutoff
localisation) uncertainty and $\delta_{\mathrm{jit},k}$ from separatrix
detection error; typical DIII-D magnitudes are $\sigma_{\mathrm{bias}}
\approx 0.02$, $\sigma_{\mathrm{jit}} \approx 0.01$ in $\rho$ units.

\section{Method}
\label{sec:method}

\paragraph{FNO surrogate.}
The forward model is a one-dimensional FNO~\cite{li2021fno} trained on 
Eq.~\eqref{eq:transport} trajectories obtained by implicit Euler solver. 
Inputs are eight channels on the $\rho$ grid: initial profile ($n_e(t=0)$),
geometry ($G_0$, $G_1$, $V'$, see ~\cite{blanken2018}), spatial coordinate ($\rho)$,
and transport/source profiles $D_{\kappa}(\rho)$, $v_{\kappa}(\rho)$,
$S_{\kappa}(\rho)$ (parametrized by $\kappa \in \mathbb{R}^{11}$);
the output is the profile at the next 64 time steps spanning 0.5\,s.
On held-out trajectories a single trained network reaches
$\approx 3.0\%$ relative $L^2$ error. One rollout evaluates in 3.0\,ms
versus 155.4\,ms for the solver (51.3$\times$ speedup), both on a single
GPU (batch size 1).

\paragraph{Training loss.}
The FNO is trained with a physics-informed loss combining a supervised
data term and a physics-residual term,
$L_{\mathrm{train}}=w_{\mathrm{data}} \cdot \text{rel-}L^2(\hat u, u) + w_{\mathrm{pde}}
\cdot \text{MSE}(R(\hat u))$, where $\hat u$ is the network's predicted trajectory,
$u$ the solver's, and $R$ is the strong-form residual of
Eq.~\eqref{eq:transport} evaluated on the network's own output via
finite differences in $t$ and $\rho$. We use $w_{\mathrm{data}} = 1$
and $w_{\mathrm{pde}} = 10^{-4}$.

\paragraph{Reconstruction loss.}
Reconstruction optimizes the transport and source parameters $\kappa$ by
minimizing a sum of squared residuals over time and observation
points,
\begin{equation}
  L(\kappa) = \frac12\sum_k \big[y_k^{\mathrm{avg}}(\kappa) - \tilde y^{\mathrm{avg}}_k\big]^2
  + \frac12\sum_{k,j} \big[y_{k,j}^{\mathrm{edge}}(\kappa) - \tilde y^{\mathrm{edge}}_{k,j}\big]^2,
\end{equation}
where $y_k^{\mathrm{avg}}(\kappa)$, $y_{k,j}^{\mathrm{edge}}(\kappa)$ are the
observation operators~\eqref{eq:obsop} applied to the predicted
profile (line average and profile evaluated at the nominal, unshifted
$\rho_j$), while $\tilde y^{\mathrm{avg}}_k$, $\tilde y^{\mathrm{edge}}_{k,j}$
are the corresponding measured values -- either real DIII-D data, or
synthetic data drawn once from the noise model~\eqref{eq:noisemodel} and
held fixed. Either way the noise ($\varepsilon_k$, $\delta_k$, or real
instrumental noise) belongs only to that fixed target, never to the
prediction. The sums run over all time steps $k$ and, for the edge term,
all observed $\rho_j \ge \rho_{\mathrm{obs}}$.

\paragraph{Posterior approximations.}
We compare four approximations of $p(\kappa \mid \tilde y)$:
\begin{itemize}
  \item \textbf{Multi-seed ensemble}: the set of reconstructed solutions
    $\{\kappa_1, \dots, \kappa_{N_{\mathrm{seed}}}\}$, a deep-ensemble
    style estimate of multi-modality.
  \item \textbf{Laplace}~\cite{mackay1992}: a Gaussian
    $\mathcal{N}(\hat\kappa, \Sigma)$ with
    $\Sigma^{-1} = H / \hat\sigma^2$, where $H$ is the Hessian
    of $L(\kappa)$ at $\hat\kappa$ from automatic differentiation
    through the surrogate. Eigenvalues are clamped at a fixed condition
    number to regularise flat directions.
  \item \textbf{Markov Chain Monte Carlo (MCMC)}: Metropolis--Hastings algorithm with isotropic
    random-walk proposals in normalized parameter space, acceptance
    $\alpha = \min\{1, \exp[-(L(\kappa') -
    L(\kappa))/\hat\sigma^2]\}$, where $\hat\sigma^2$ is a
    tuned temperature.
  \item \textbf{Gaussian method}: an isotropic Gaussian $\mathcal{N}(\hat\kappa, \sigma)$, 
    independent of the reconstruction fit, with $\sigma$
    calibrated once against held-out discharges (\S\ref{sec:experiments}).
\end{itemize}
Each approximation yields parameter samples; pushing them through the
forward model gives an ensemble of profile evolutions from which we
compute predictive bands.

\paragraph{Physics-free baseline.}
As the representative of current practice we fit a modified hyperbolic
tangent profile~\cite{groebner1998} to each time slice independently,
given exactly the same measured values $\tilde y$: mtanh
parameters minimize the same objective $L(\kappa)$ above, via the same Adam
procedure (Appendix~\ref{app:repro}). This baseline has no temporal or transport coupling.

\paragraph{Metrics.}
Point accuracy is reported separately on the unobserved axis ($\rho < 0.05$), 
core ($\rho < 0.7$) and the full profile as the relative $L^2$ error 
of the reconstruction against the ground-truth (synthetic) or measured (DIII-D) profile evolution. 
Predicted uncertainty is ranked by the
CRPS~\cite{gneiting2007}, a scoring rule that for an ensemble
forecast reduces to
$\mathrm{CRPS} = \mathbb{E}|X - y| - \tfrac{1}{2}\mathbb{E}|X - X'|$,
where $X, X'$ are independent draws from the predictive ensemble and
$y$ is the truth: the first term rewards proximity to the truth, the
second penalizes an unnecessarily wide (undersharp) ensemble, so a
lower score jointly rewards calibration and sharpness without
requiring a parametric predictive distribution. We additionally report the
empirical coverage of the exact
1$\sigma$/2$\sigma$ central predictive intervals (0.6827 / 0.9545).

\section{Experiments}
\label{sec:experiments}

\paragraph{Setup.}
Synthetic evaluation dataset contains 100 discharges generated by the implicit-Euler solver
with transport and source parameters drawn from the prior and observation noise 
$\sigma$.
DIII-D dataset contains 13 discharges (L- and H-mode), with density
profiles reconstructed from Thomson scattering measurements.

\begin{figure}[h]
  \centering
  \includegraphics[width=0.48\textwidth]{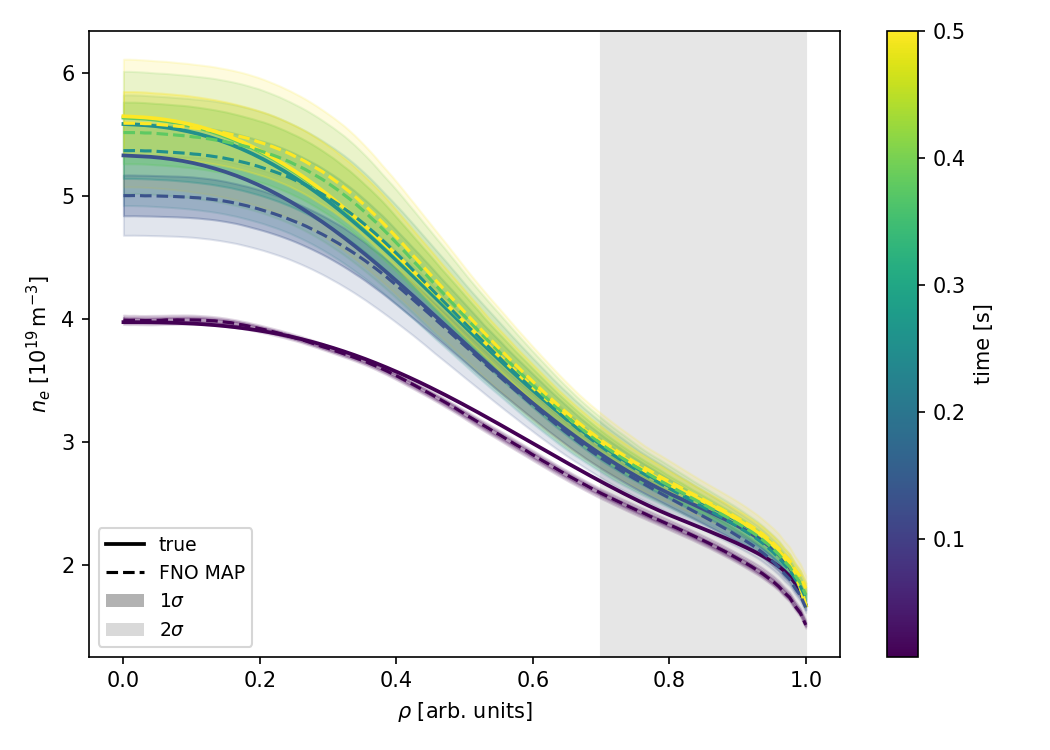}
  \includegraphics[width=0.48\textwidth]{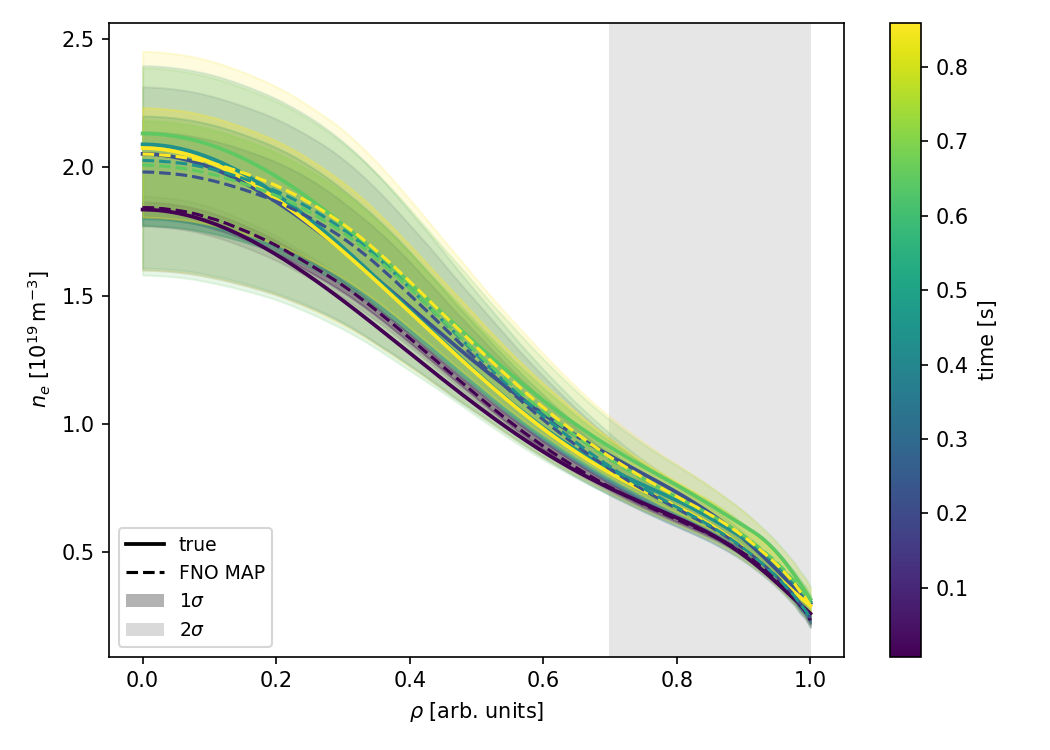}
  \caption{Reconstruction of the unobserved core over time, for 
    synthetic discharge (left) and DIII-D discharge (right). 
    Solid: true profile; dashed: FNO MAP reconstruction;
    shaded bands: 1$\sigma$/2$\sigma$ posterior spread (multi-seed
    ensemble); color encodes time within the reconstructed window.
    Shaded $\rho$ region: observed edge ($\rho \ge 0.7$).}
  \label{fig:reconstruction}
\end{figure}

\paragraph{Reconstruction accuracy.} Relative $L^2$ error of density profile reconstruction is shown in Fig.~\ref{fig:reconstruction}
and Table~\ref{tab:reco}. On both the synthetic and experimental datasets, the transport-based FNO model
outperforms the mtanh baseline at the magnetic axis by a significant margin: $0.057$ against mtanh's $0.080$ ($29\%$ lower) on the synthetic dataset, and
$0.103$ against $0.161$ ($36\%$ lower) on DIII-D discharges. At the core and full-profile regions FNO underperforms mtanh instead
(by $8$--$15\%$ on synthetic, $13$--$23\%$ on DIII-D) since mtanh fits $5$ parameters per time slice (roughly $320$ total over a typical
reconstruction window) directly against the observed edge and line-integrated interferometer signal, against FNO's $11$
transport/source parameters shared across the whole discharge -- far more flexibility to fit the data, at the cost of
any physical constraint on the unobserved core.

\begin{table}[h]
  \caption{MAP relative $L^2$ error, for
    the realistic-noise synthetic dataset with hidden plasma core (100 discharges)
    and for 13 experimental DIII-D discharges.
    core = $\rho<0.7$ (unobserved); axis = $\rho<0.05$ (unobserved).}
  \label{tab:reco}
  \centering
  \begin{tabular*}{\textwidth}{@{\extracolsep{\fill}}lccc}
    \toprule
    Method & axis & core & full \\
    \midrule
    \multicolumn{4}{l}{\textit{Synthetic}} \\
    FNO   & $\mathbf{0.0565 \pm 0.0450}$ & $0.0596 \pm 0.0334$ & $0.0657 \pm 0.0353$ \\
    mtanh & $0.080 \pm 0.083$ & $\mathbf{0.055 \pm 0.051}$ & $\mathbf{0.056 \pm 0.058}$ \\
    \midrule
    \multicolumn{4}{l}{\textit{DIII-D}} \\
    FNO   & $\mathbf{0.103 \pm 0.076}$ & $0.102 \pm 0.035$ & $0.106 \pm 0.033$ \\
    mtanh & $0.161 \pm 0.097$ & $\mathbf{0.089 \pm 0.054}$ & $\mathbf{0.082 \pm 0.053}$ \\
    \bottomrule
  \end{tabular*}
\end{table}

Fig.~\ref{fig:reconstruction} shows the reconstructed profile's time evolution with uncertainty bands, confirming the
transport-based model identifies coefficients that reproduce the observed behaviour on both synthetic and
experimental discharges.

\paragraph{Uncertainty quantification.}
We calibrate each method's free parameter(s) against 10 held-out synthetic trajectories using
coverage at nominal level $\alpha$, the empirical fraction of held-out truth values falling inside
the ensemble's $\alpha$-central interval, e.g.\ between the $(1-\alpha)/2$ and $(1+\alpha)/2$
quantiles. A well-calibrated method has coverage equal to $\alpha$; below $\alpha$ is overconfident
(too narrow), above is underconfident (too wide). Gaussian method's $\sigma$, MCMC's proposal step size $\sigma$ and
Laplace's noise variance $\hat\sigma^2$ and Hessian clamp ratio are all calibrated by directly
minimizing mean calibration error $|\mathrm{coverage}(\alpha) - \alpha|$ averaged over
$\alpha \in [0.5, 0.95]$ with Optuna. MCMC's chain temperature self-calibrates per shot as
$\hat\sigma^2_{\mathrm{MCMC}} = L(\hat\kappa)$ (the MAP loss), independently of the calibrated
proposal step size, so a poorer-fitting shot automatically samples a more diffuse chain.

Table~\ref{tab:uq} reports CRPS and $1\sigma$/$2\sigma$ coverage
for FNO's four UQ methods. On the synthetic dataset, the Gaussian method achieves the best CRPS and the coverage
closest to nominal in every column; multi-seed, Laplace, and MCMC are all underconfident in the unobserved core at
$1\sigma$, and the gap widens toward the axis. On the DIII-D discharges the pattern is more mixed: the Gaussian
method still wins CRPS in both regions and edge coverage at $1\sigma$, but Laplace is closest to nominal for core
coverage and both regions, despite it was tunned only on synthetic data.

\begin{figure}[t]
  \centering
   \includegraphics[width=\textwidth]{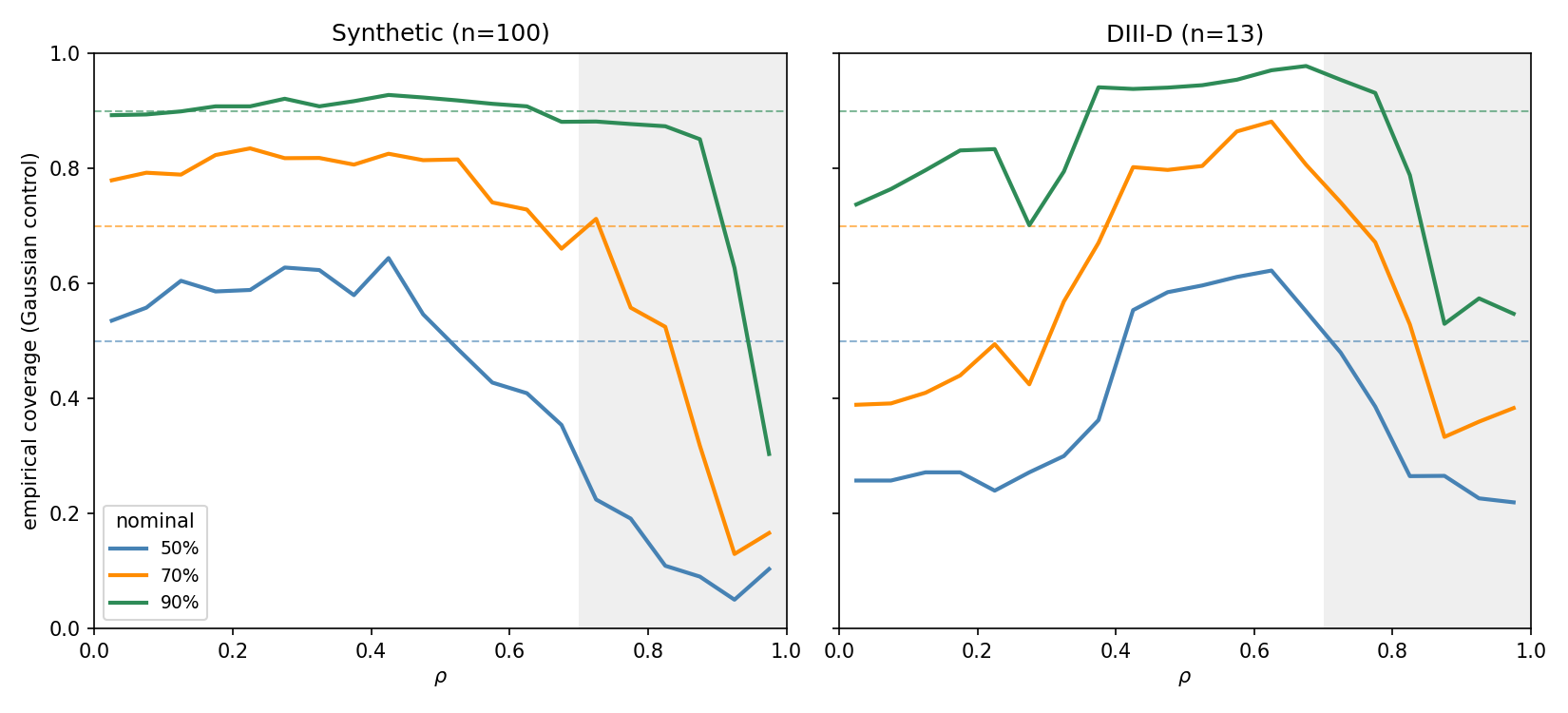}
  \caption{Gaussian-method empirical vs.\ nominal coverage at the
    50/70/90\% levels, as a function of $\rho$ (median per level, binned
    into 20 $\rho$ intervals; shaded region marks the observed edge,
    $\rho \ge 0.7$), comparing the 100 synthetic discharges (left)
    against the 13 DIII-D discharges (right).
    IQR bands are omitted for readability -- per-shot spread is wide and
    reported numerically as the $\pm$ std in Table~\ref{tab:uq}.}
  \label{fig:calibration}
\end{figure}

Fig.~\ref{fig:calibration} shows the Gaussian method's calibration profile across $\rho$: at the
50/70/90\% nominal levels, coverage on the synthetic sweep tracks its nominal level near the axis but becomes
increasingly overconfident (coverage well below nominal) past the observed edge boundary at $\rho=0.7$; on the DIII-D
discharges the pattern inverts near the axis (overconfident there too), with better-calibrated coverage through the
mid-profile before falling again at the deepest edge. The Gaussian method's calibration is thus not uniform in $\rho$,
nor does it transfer as-is from synthetic to experimental data.

\begin{table}[t]
  \caption{Reconstruction and UQ quality, median $\pm$ std over
    100 synthetic discharges (top) and 13 DIII-D
    discharges (bottom), FNO only. Core = unobserved $\rho < 0.7$; axis =
    $\rho < 0.05$, the deepest extrapolation from the observed window;
    edge = observed $\rho \ge 0.7$. Best per column in bold. Both MCMC
    and Laplace are underconfident in the unobserved core at $1\sigma$.
    Reconstruction accuracy (rel.\ err) is identical across UQ methods
    within each model (shared MAP fit) and is reported in
    Table~\ref{tab:reco}.}
  \label{tab:uq}
  \centering
  \scriptsize
  \setlength{\tabcolsep}{3pt}
  \begin{tabular}{lcccccc}
    \toprule
    & \multicolumn{2}{c}{CRPS} & \multicolumn{2}{c}{cov.\ $1\sigma$ (0.68)} & \multicolumn{2}{c}{cov.\ $2\sigma$ (0.95)} \\
    \cmidrule(lr){2-3} \cmidrule(lr){4-5} \cmidrule(lr){6-7}
    Method & core & edge & core & edge & core & edge \\
    \midrule
    \multicolumn{7}{l}{\textit{FNO} ($n=100$)} \\
    Multi-seed        & $0.087 \pm 0.108$ & $0.091 \pm 0.119$ & $0.235 \pm 0.258$ & $0.212 \pm 0.304$ & $0.398 \pm 0.296$ & $0.334 \pm 0.341$ \\
    Laplace           & $0.091 \pm 0.110$ & $0.082 \pm 0.124$ & $0.262 \pm 0.274$ & $0.244 \pm 0.298$ & $0.561 \pm 0.312$ & $0.619 \pm 0.354$ \\
    MCMC              & $0.101 \pm 0.109$ & $0.105 \pm 0.158$ & $0.239 \pm 0.181$ & $0.128 \pm 0.232$ & $0.467 \pm 0.273$ & $0.259 \pm 0.324$ \\
    Gaussian control  & $\mathbf{0.082 \pm 0.095}$ & $\mathbf{0.075 \pm 0.116}$ & $\mathbf{0.660 \pm 0.281}$ & $\mathbf{0.336 \pm 0.303}$ & $\mathbf{0.921 \pm 0.220}$ & $\mathbf{0.780 \pm 0.307}$ \\
    \midrule
    \multicolumn{7}{l}{\textit{FNO, DIII-D discharges} ($n=13$)} \\
    Multi-seed        & $0.205 \pm 0.231$ & $0.196 \pm 0.231$ & $0.115 \pm 0.063$ & $0.067 \pm 0.019$ & $0.188 \pm 0.092$ & $0.102 \pm 0.032$ \\
    Laplace           & $0.324 \pm 0.294$ & $0.381 \pm 0.290$ & $\mathbf{0.547 \pm 0.320}$ & $0.405 \pm 0.355$ & $\mathbf{0.969 \pm 0.255}$ & $\mathbf{0.951 \pm 0.348}$ \\
    MCMC              & $0.339 \pm 0.244$ & $0.207 \pm 0.290$ & $0.369 \pm 0.308$ & $0.292 \pm 0.108$ & $0.501 \pm 0.314$ & $0.556 \pm 0.161$ \\
    Gaussian control  & $\mathbf{0.154 \pm 0.206}$ & $\mathbf{0.165 \pm 0.216}$ & $0.495 \pm 0.274$ & $\mathbf{0.434 \pm 0.276}$ & $0.871 \pm 0.215$ & $0.798 \pm 0.263$ \\
    \bottomrule
  \end{tabular}
\end{table}


\section{Conclusion and Limitations}
\label{sec:limitations}

We presented a differentiable FNO surrogate for transport-based reconstruction of the unobserved plasma core, with
four posterior approximations compared against a physics-free mtanh baseline on synthetic and 13 DIII-D discharges.
The transport, source, and geometry models are fixed, low-dimensional parametric forms (piecewise-linear
diffusivity/pinch, exponential wall-fuelling plus Gaussian core source, static EFIT-derived geometry); a true
profile shaped outside these families cannot be represented regardless of how well $\kappa$ is inferred, and
time-varying geometry within a discharge is not modelled. Ongoing work extends the forward model to coupled density
and temperature transport with a Bohm/gyro-Bohm closure~\cite{erba1998} for inference of its dimensionless
coefficients on DIII-D, and compares the surrogate against direct differentiation of the implicit solver.

\newpage


\bibliographystyle{unsrtnat}
\bibliography{references}


\appendix

\section{Reproducibility details}
\label{app:repro}

\paragraph{Solver and training data.}
Training trajectories are generated with an implicit-Euler,
LU-factored solve of Eq.~\eqref{eq:transport} on a grid of $128$
$\rho$-points $\times$ $64$ time steps per $0.5$\,s segment. $1000$
training, $200$ validation, and $200$ test trajectories are drawn by
sampling transport/source coefficients $\kappa$ and geometry
coefficients uniformly from the prior ranges in Table~\ref{tab:priors};
initial conditions are sampled from a pool of EFIT reconstructed
real DIII-D $n_e(\rho)$ shots.

\begin{table}[h]
  \caption{Prior ranges for the transport/source coefficients $\kappa$.}
  \label{tab:priors}
  \centering
  \small
  \begin{tabular}{llc}
    \toprule
    Group & Parameter & Range \\
    \midrule
    \multirow{4}{*}{diffusion} & $\rho_{\mathrm{ped}}$ & $[0.7, 0.9]$ \\
     & $D_{\mathrm{core}}$ & $[0.3, 3.0]$ \\
     & $D_{\mathrm{ped}}$ & $[0.3, 2.0]$ \\
     & $D_{\mathrm{edge}}$ & $[0.05, 0.5]$ \\
    \midrule
    \multirow{3}{*}{pinch} & $v_{\mathrm{core}}$ & $0$ (fixed) \\
     & $v_{\mathrm{ped}}$ & $[-0.5, 0.5]$ \\
     & $v_{\mathrm{edge}}$ & $[-2.0, 2.0]$ \\
    \midrule
    \multirow{2}{*}{wall source} & amplitude $A_w$ & $[0.0, 500.0]$ \\
     & decay length $\lambda$ & $[0.002, 0.08]$ \\
    \midrule
    \multirow{3}{*}{core source} & amplitude $A_c$ & $[0.0, 50.0]$ \\
     & center $\rho_0$ & $[0.0, 0.5]$ \\
     & width $\sigma_0$ & $[0.01, 0.5]$ \\
    \bottomrule
  \end{tabular}
\end{table}

\paragraph{FNO architecture and training.}
$16$ Fourier modes, $128$ hidden channels, $4$ layers, projection
channel ratio $2$. Trained for $1000$ epochs, batch size $32$, Adam
with $\mathrm{lr}=5\times10^{-3}$ and weight decay $10^{-4}$, one-cycle
learning-rate schedule (max lr $10^{-3}$, $5\%$ warmup, div factor
$10$, final div factor $1000$). Loss weights $w_{\mathrm{data}}=1$,
$w_{\mathrm{pde}}=10^{-4}$ (Eq. in \S\ref{sec:method}). Training
completes in $\approx 13$ minutes on a single GPU.

\paragraph{Reconstruction optimizer.}
MAP fitting uses Adam with $\mathrm{lr}=0.05$ and $5$ random restarts
(the best of which is kept as $\hat\kappa$). FNO converges in $500$
steps; mtanh's per-slice fit ($5$ parameters $\times$ $64$ time steps
$\approx 320$ free parameters for a single $0.5$\,s segment) needs
$5000$ steps to converge.

\paragraph{UQ settings.}
Multi-seed method estimates uncertainty from the $5$ restart solutions directly. 
The Gaussian method draws $500$ samples with an isotropic width $\sigma=0.045$
(4.5\% of the prior range). MCMC runs $5000$
Metropolis--Hastings steps, retaining $500$ samples, with the chain
temperature self-calibrating per shot ($\hat\sigma^2_{\mathrm{MCMC}} =
L(\hat\kappa)$), with the proposal step size $\sigma=3.03\times10^{-3}$.
Laplace draws $500$ samples from a Hessian-based Gaussian with
eigenvalues floored at a clamp ratio of the largest eigenvalue $\mathrm{ratio}=2.87\times10^{-6}$ and noise variance $\hat\sigma^2=99.45$. 
The Gaussian, MCMC, and Laplace parameters above were calibrated with
Optuna by directly minimizing mean calibration error on the same
held-out trajectories.

\paragraph{Hardware and runtimes.}
All training and reconstruction runs on a single GPU. FNO training:
$\approx 13$ minutes for $1000$ epochs. A single FNO rollout: $3.0$\,ms
versus $155.4$\,ms for the implicit-Euler solver ($51.3\times$
speedup, batch size $1$). Full reconstruction (MAP fit plus all four
UQ methods) per DIII-D discharge: $\approx 2.5$ minutes for FNO,
$\approx 2.5$ minutes for mtanh (MCMC's $5000$ sampling steps dominate
regardless of the forward model's own cost). The full sweep reported
in \S\ref{sec:experiments} (100 synthetic + 13 DIII-D discharges, both
models) took $\approx 2.5$ hours (FNO) and $\approx 4$ hours (mtanh)
wall-clock.

\end{document}